# On the Inference of the Cosmic Ray Ionization Rate $\zeta$ from the HCO$^+$-to-DCO$^+$ Abundance Ratio: The Effect of Nuclear Spin

Christopher N. Shingledecker

*Department of Chemistry, University of Virginia, Charlottesville, VA 22904*

`shingledecker@virginia.edu`

Jennifer B. Bergner

*Department of Chemistry and Chemical Biology, Harvard University, Cambridge, MA 02138*

Romane Le Gal

*Department of Chemistry, University of Virginia, Charlottesville, VA 22904*

Karin I. Öberg

*Harvard-Smithsonian Center for Astrophysics, 60 Garden Street, Cambridge, MA 02138*

Ugo Hincelin

*Department of Chemistry, University of Virginia, Charlottesville, VA 22904*

and

Eric Herbst

*Department of Chemistry, University of Virginia, Charlottesville, VA 22904*

*Department of Astronomy, University of Virginia, Charlottesville, VA 22904*

## ABSTRACT

The chemistry of dense interstellar regions was analyzed using a time-dependent gas-grain astrochemical simulation and a new chemical network that incorporates deuterated chemistry taking into account nuclear spin-states for the hydrogen chemistry and its deuterated isotopologues. With this new network, the utility of the [HCO$^+$]/[DCO$^+$] abundance ratio as a probe of the cosmic ray ionization rate has been reexamined, with special attention paid to the effect of the initial value of the molecular hydrogen ortho-to-para ratio (OPR). After discussing the use of the probe for cold cores, we then compare our results with previous theoretical and observational results for a molecular cloud close to the supernova remnant W51C, which is thought to have an enhanced cosmic ray ionization rate $\zeta$ caused by the nearby $\gamma$-ray source. In addition, we attempt to use our approach to estimate the cosmic ray ionization rate for L1174, a dense core with an embedded star. Beyond the previously known sensitivity of $[HCO^+]/[DCO^+]$ to $\zeta$, we demonstrate its additional dependence on the initial OPR and, secondarily, on the age of the source, its temperature, and its density. We conclude that the usefulness of the $[HCO^+]/[DCO^+]$ abundance ratio to constrain the cosmic ray ionization rate in dense regions increases with source age and ionization rate as the ratio becomes far less sensitive to the initial value of the OPR.

## 1. Introduction

In dense interstellar clouds, cosmic rays and their associated secondary electrons act as the dominant ionization source, since most other types of radiation such as external UV photons are quickly attenuated. The cosmic ray ionization rate, $\zeta$, is generally taken to be $\sim 10^{-17}$ s$^{-1}$ (Spitzer & Tomasko 1968; Indriolo & McCall 2013), though this value is likely enhanced in certain sources such as diffuse clouds, outer regions of dense clouds, and supernova remnants (McCall et al. 2002; Rimmer et al. 2012; Ceccarelli et al. 2011). This value is important to constrain because it sets the ion fraction of dense clouds and will therefore have a significant impact on the chemistry. Given the astrochemical importance of ion-neutral reactions, cosmic rays are the drivers of a type of chemistry that is particularly efficient, even in cold dense regions. It is therefore reasonable to expect that regions of the interstellar medium (ISM) with different cosmic ray ionization rates will exhibit different gas-phase chemical abundances (Farquhar et al. 1994).

Indeed, the cosmic ray ionization rate can be probed by observations of specific chemical abundances. The first estimates of the cosmic ray ionization rate were based on observations of OH and HD in diffuse clouds, assuming that both molecules are formed starting with charge exchange reactions involving $H^+$ and, respectively, O and D, with the proton resulting from ionization of atomic hydrogen by cosmic rays (O'Donnell & Watson 1974; Black & Dalgarno 1977; Black et al. 1978; Hartquist et al. 1978). However, other physical and chemical processes complicate this procedure and can introduce substantial uncertainties. For instance, competing pathways for $H^+$ destruction by polycyclic aromatic hydrocarbons (PAHs) and small grains have been studied in the context of diffuse clouds (Wolfire et al. 2003; Liszt 2003). Effects caused by magnetic fields, which can impact the ionization rate gradient in sources due to the enhanced attenuation of cosmic rays have also been examined in dense cloud models (Rimmer et al. 2012; Padovani et al. 2009). Furthermore, the chemistry of deuterium and oxygen has been deepened and revised, leading to the conclusion that the main formation pathways for HD and OH molecules depend on the sources in which they are found. For instance, in dense gas OH is formed mainly through a sequence of reactions starting with $O + H_3^+$ while HD is formed mainly on dust surfaces (Mennella 2008; Manicò et al. 2001). For diffuse sources, the detection of $H_3^+$ (Geballe & Oka 1996) gave access to a more direct, and thus more reliable, probe for the cosmic ray ionization rate due to its simple straightforward chemistry (Indriolo & McCall 2013).

An alternative approach specific to dense clouds that has been proposed uses observations of deuterated molecules to determine deuterium fractionation and, in turn, the ionization fractionation of an interstellar object (Guelin et al. 1977, 1982; Wootten et al. 1979; Dalgarno & Lepp 1984; Caselli et al. 1998; Williams et al. 1998). Specifically, for a steady-state gas-phase system, the ratio $[\text{HCO}^+]/[\text{DCO}^+]$ is proportional to the ion fraction, as described by Caselli (2002), so this ratio is a potential probe for the cosmic ray ionization rate in such gas. These molecules proceed from the reaction of CO with $\text{H}_3^+$ and $\text{H}_2\text{D}^+$ as follows:

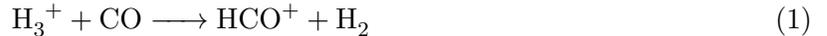

$$\text{H}_3^+ + \text{CO} \longrightarrow \text{HCO}^+ + \text{H}_2 \tag{1}$$

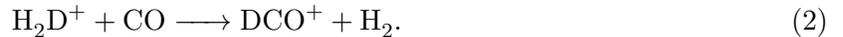

$$\text{H}_2\text{D}^+ + \text{CO} \longrightarrow \text{DCO}^+ + \text{H}_2. \tag{2}$$

CO is of prime importance here since it constitutes the dominant neutral species that can accept a proton from $\text{H}_3^+$.

The abundance ratio of $[\text{H}_3^+]/[\text{H}_2\text{D}^+]$ is known to be reduced in cold regions due to the slight exothermicity of a proton transfer from $\text{H}_3^+$ to HD:

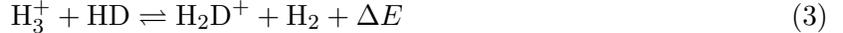

$$\text{H}_3^+ + \text{HD} \rightleftharpoons \text{H}_2\text{D}^+ + \text{H}_2 + \Delta E \tag{3}$$

with $\Delta E/k = 232$ K, ignoring nuclear spin (Roberts et al. 2002). So, at the low temperatures prevailing in dense clouds, the reverse reaction is inhibited, which provides an enhancement of the abundances of deuterated species and reduces the $[\text{H}_3^+]/[\text{H}_2\text{D}^+]$ ratio. $\text{H}_2\text{D}^+$ and $\text{H}_3^+$ are both destroyed by proton transfer to a neutral species, as in reactions (1) and (2), or by dissociative recombination. The importance of the proton transfer reactions shows that the initial deuterium fractionation can be propagated in species that are formed by reaction with $\text{H}_3^+$.

The main destruction pathway of both $\text{HCO}^+$ and $\text{DCO}^+$ occurs via dissociative recombination:

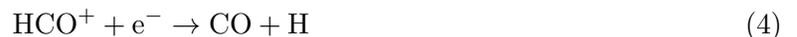

$$\text{HCO}^+ + \text{e}^- \rightarrow \text{CO} + \text{H} \tag{4}$$

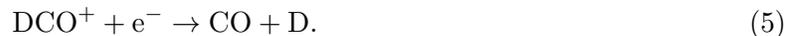

$$\text{DCO}^+ + \text{e}^- \rightarrow \text{CO} + \text{D}. \tag{5}$$

The abundance ratio, $R_\text{D}$, of $\text{HCO}^+$ to $\text{DCO}^+$ can be approximated in terms of a few key reactions at steady-state. This formalism assumes that $\text{H}_2\text{D}^+$ is formed mainly via reaction (3), and destroyed via the backwards reaction of (3), as well as by reactions with CO, $\text{e}^-$, and HD, the last of which leads to the doubly deuterated species $\text{D}_2\text{H}^+$ (Guelin et al. 1977; Caselli 2002; Furuya et al. 2015). Ignoring nuclear spins, $R_\text{D}$ is given by the equation

$$R_{\mathrm{D}} = \frac{n(\mathrm{HCO^+})}{n(\mathrm{DCO^+})} = 3 \left( \frac{n(\mathrm{H_3^+})}{n(\mathrm{H_2D^+})} \right)$$
$$= 3 \left( \frac{k_2 \, n(\mathrm{CO}) + k_{H_2D^+-e} \, n(e) + k_{HD} \, n(\mathrm{HD}) + k_{3b} \, n(\mathrm{H_2})}{k_{3f} \, n(\mathrm{HD})} \right) \tag{6}$$

where $k_{3f}$ and $k_{3b}$ are the forward and backwards rate coefficients for reaction (3), $k_2$ is the rate coefficient for reaction (2), $k_{H_2D^+-e}$ is the rate coefficient for the dissociative recombination of $\mathrm{H_2D^+}$, and $k_{HD}$ is the rate coefficient for the reaction between $\mathrm{H_2D^+}$ and HD. Equation (6) contains several assumptions that limit its applicability by affecting the relationship between $\zeta$ and $R_{\mathrm{D}}$ (Caselli 2002). Thus, this formalism is mainly useful for understanding the key chemical dependencies of $R_{\mathrm{D}}$ and is only in agreement with the full model at steady-state. Generally, the [HCO$^+$]/[DCO$^+$] ratio in any given environment may depend on a combination of 1) the ionization rate, $\zeta$ (Guelin et al. 1982; Wootten et al. 1979; Ceccarelli et al. 2011; Caselli 2002) 2) CO freeze-out (Caselli 2002), 3) gas temperature (Caselli et al. 1998; Ceccarelli et al. 2011), 4) chemical time (Caselli 2002), 5) formation pathways (Dalgarno & Lepp 1984; Caselli 2002), and 6) nuclear spin state populations.

This last complicating factor concerns the different chemical properties of ortho and para states of certain species, which are thought to play an important role in interstellar deuterium fractionation (Pagani et al. 2009). In the case of $\mathrm{H_2}$, the para form, in which the nuclear spins are anti-parallel, is about 170 K lower in energy than the ortho form, in which the spins are parallel. The fact that ortho hydrogen is higher in energy than para hydrogen is important in the context of deuteration, since, depending on the spin states of $\mathrm{H_3^+}$, $\mathrm{H_2D^+}$, and $\mathrm{H_2}$ in reaction (3), the backwards reaction can be exothermic by 25 K if both $\mathrm{H_2D^+}$ and $\mathrm{H_2}$ are in the ortho state and $\mathrm{H_3^+}$ is para, or have a reduced barrier for other combinations of spin states (Hugo et al. 2009; Furuya et al. 2015). As derived in Furuya et al. (2015), the rate coefficient for the backwards reaction can be expressed as a function of the ortho-to-para ratio for $\mathrm{H_2}$, often abbreviated as OPR, and the gas temperature; however, the OPR of $\mathrm{H_2}$ in cold dense clouds is not observationally constrained to the best of our knowledge. Theory suggests that this OPR is very small and time independent at late times, but the initial $\mathrm{H_2}$ OPR is unknown. If we make the assumption that diffuse clouds represent the initial stages of dense clouds, at least as regards molecular hydrogen, we can use the large number of measurements of this OPR in diffuse clouds, which were made some time ago by Savage et al. (1977). The majority of the measurements, based on electronic transitions starting with J=0 and J=1 in the ground electronic-vibrational state, yield OPR results in the range 0.5-1.5, although clouds with $\mathrm{H_2}$ OPRs over a wider range were detected.

The underlying chemistry behind the observed variability in [HCO$^+$]/[DCO$^+$] with initial OPR is fairly straightforward. The higher the initial OPR of a source, the higher the rate of the backward reaction of (3) due to an increased abundance of *ortho*-hydrogen. This increase reduces the amount of $\mathrm{H_2D^+}$, the initiator for subsequent deuteration reactions (Pagani et al. 2009). A higher OPR will thus result in fewer deuterated species which, in turn, will lead to a higher [HCO$^+$]/[DCO$^+$] ratio.

Over time, the OPR of a source will gradually evolve towards a steady-state value (*e.g.* Le Gal et al. 2014, and references therein) driven mainly by reactions between the ions $H^+$ and $H_3^+$ with $H_2$. Eventually, differences in the initial OPR for systems with the same value of $\zeta$ will disappear at rates depending upon the cosmic ray ionization rate, and $\zeta$ will correlate more strongly with the abundance ratio. Specifically, increasing $\zeta$ has the effect of reducing the time it takes for the OPR of a source to reach its steady-state value, since rates for the production of $H^+$ and $H_3^+$ increase as well. The nuclear spin effect can result in a breakdown of the relation between the $[HCO^+]/[DCO^+]$ abundance ratio and the cosmic ray ionization rate, before the OPR reaches a steady-state value. This effect is discussed further in Section 3.1.

Here, we present the results of our attempt to elucidate the additional effects in the models caused by an explicit consideration of nuclear-spin chemistry, as well as some additional complications caused by uncertainties in source lifetime, temperature, and density. The model results are first compared in general with previous results of Caselli et al. (1998), who simulated a number of dense cores in which $[HCO^+]/[DCO^+]$ had been measured. We then consider two specific sources: (i) a molecular cloud close to the supernova remnant W51C, already studied by Ceccarelli et al. (2011) who made an attempt to infer the cosmic ray ionization rate from this abundance ratio, and (ii) the cloud L1174, one of the cores discussed in Caselli et al. (1998), a dense core with an embedded star.

The organization of the rest of the paper is as follows. In Section 2 descriptions of the model, chemical network, and sensitivity analyses are given, while in Section 3, the results of our investigation are presented. Section 4 contains a discussion of the results while Section 5 presents our conclusions.

## 2. Model

In order to understand the effect of nuclear spin on the chemistry of relevant environments, three sets of physical conditions were used. First, $HCO^+$ and $DCO^+$ abundances were investigated under cold core conditions with a gas temperature of 10 K and a density of $10^4$ cm$^{-3}$. The second set of physical properties reflects those of the molecular cloud close to the SNR, W51C (Kundu & Velusamy 1967), which has a temperature of 24 K and density of $10^4$ cm$^{-3}$ (Ceccarelli et al. 2011). Finally, we model L1174, a dense core with an embedded protostar previously studied by Caselli et al. (1998), using a temperature of 15 K and density of $10^4$ cm$^{-3}$. For all sets of physical conditions, the interstellar environments were approximated using a "0D" homogeneous physical structure. Since the first environment considered is a generic cold core, and there is no detailed structural information available for the specific sources, a more detailed approach, which takes into account gradients for properties such as temperature and cosmic ray ionization rate (Rimmer et al. 2012; Padovani et al. 2009) was not carried out.

The `Nautilus` program (Semenov et al. 2010; Reboussin et al. 2014), a two-phase gas-grain

astrochemical code utilizing a rate-equation approach, was used for modeling the chemistry in this study. In this model, cosmic rays cause ionization due to direct impact and the impact of secondary electrons with overall rate coefficient $\zeta$, and cause photodissociation due to secondary UV photons produced by excitation of $H_2$. The rate coefficients for photodissociation are given by $k_{CR} = A_i\zeta$, where $A_i$ is a value for species $i$, as compiled by Wakelam et al. (2012). In Nautilus, grain heating caused by cosmic rays is approximated using the formalism of Hasegawa & Herbst (1993) and Leger et al. (1985). A true stochastic treatment can be found in Herbst & Cuppen (2006), although it is not useful for a complete network. Nautilus was used here along with a novel chemical network, recently developed by U. Hincelin (unpublished), which includes the nuclear spin states of key species along with deuterated species. Reactive desorption was included for grain surface reactions with the parameter $a_{RRK} = 0.01$ following the technique used by Garrod et al. (2007).

The elemental abundances used in our calculations are the so-called low-metal abundances (Graedel et al. 1982; Lee et al. 1996). Caselli et al. (1998) and Maret & Bergin (2007) investigated the impact of metallicity on the relationship between $\zeta$ and $R_D$ and found that changes in the abundance of refractory metals (e.g. Fe, Mg, and Na) can have an effect on the value of $[HCO^+]/[DCO^+]$ mainly through influencing electron abundances in a region. We examined this in our network by varying the abundance of $M^+$, where $x(M^+) = x(Mg^+) + x(Fe^+) + x(Na^+)$, between 0.1 and 10 times the standard value of $1.2 \times 10^{-8}$ used in Lee et al. (1996). From the steady-state results of these models, we determined that there was little to no deviation of the result for $R_D$ within the range of metallicities tested for models in which the standard ionization rate of $\zeta = 1.3 \times 10^{-17}$ s$^{-1}$ was used, and those using an enhanced value of $\zeta = 10^{-15}$ s$^{-1}$. Furthermore, when sulfur was included in $M^+$ and metal abundances were varied by an order of magnitude, little or no change in the steady state abundance of $R_D$ was also noted. These results are due both to the gas-phase depletion of the refractory metals at late times due to freeze out onto grains, and a relative increase in the number of gas-phase molecular ions such as $H_3O^+$ and $H_2CN^+$ which are more abundant at steady state in our code than gas-phase metals and are, thus, more important in determining electron abundances. The simulations described in this work also contain an initial fractional abundance for $H_2$ with respect to total hydrogen of $x(H_2)_{tot} = 0.499975$. This value is the sum of the fractional abundances of *ortho*-$H_2$ and *para*-$H_2$.

Sensitivity analyses were performed by running a large number of simulations with varying values for certain parameters, principally the cosmic ray ionization rate, $\zeta$, and the initial value of the $H_2$ OPR, in order to determine their impact on the simulations (Wakelam et al. 2010). For this work, $\zeta$ was varied between $10^{-18}$ and $10^{-15}$ s$^{-1}$ and the initial OPR between 0.01 and 3. The selection for the range of initial OPR values should be seen as a range of possible values (Furuya et al. 2015; Vaupré et al. 2014), with the range used for the cosmic ray ionization rate likewise merely representing a set of values that spans a region emcompassing most of the estimates for $\zeta$ inferred from observations and theory (Rimmer et al. 2012). Unless otherwise noted, in cases where only $\zeta$ was varied, an initial OPR value of 0.01 was assumed (Troscompt et al. 2009; Dislaire et al. 2012; Vaupré et al. 2014). In addition to varying $\zeta$ and the initial OPR, we furthermore explored the

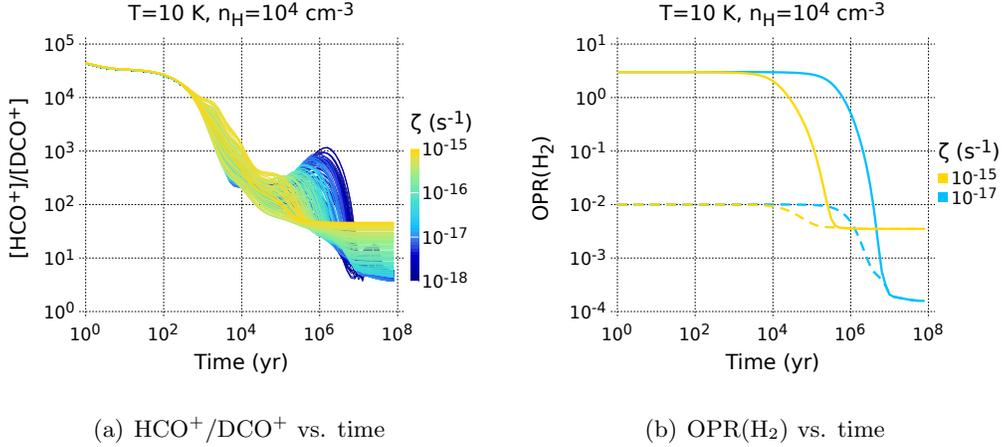

(a) HCO⁺/DCO⁺ vs. time      (b) OPR(H₂) vs. time

Fig. 1.— Panel (a): Sensitivity analysis results for [HCO+]/[DCO+] vs time in a cold core where both zeta and the initial OPR are varied randomly. Panel (b): Plot of OPR vs time for two extreme values of the initial OPR (0.01, dashed lines and 3.0, solid lines) and two values of zeta, showing that steady-state is reached more quickly with the larger value of zeta. At steady-state, the initial value of OPR is no longer of importance.

effects of the physical conditions of an interstellar region by running sensitivity analyses at densities of $10^3$ and $10^5$ cm$^{-3}$ and a kinetic temperature of 10 K. These conditions are representative of cold cores.

When either $\zeta$ or the initial OPR was varied, $y$, the specific value used for each of the simulation runs in the sensitivity analysis was sampled from a boxcar, or rectangular, probability density function between $y_{min}$ and $y_{max}$. The value of each $y$ was calculated using the expression

$$y = 10^{R_n \times \log_{10}(y_{\min}) + (1-R_n) \times \log_{10}(y_{\max})} \qquad (7)$$

where $R_n$ is a uniform pseudorandom number in the range $[0, 1]$.

## 3. Results

### 3.1. Cold Cores

Initially, simulations of a representative cold core were carried out with the dual variation of $\zeta$ and the initial OPR. The results of these models are shown in Fig. 1(a), in which [HCO⁺]/[DCO⁺] is plotted against time. From these data, one can distinguish three distinct regions that arise as $\zeta$ and the initial OPR were varied. The first can be seen between $10^0$ and $10^3$ yr in which there is no significant effect on the abundance ratio due to parameter variations. From $10^3$ to $\sim 10^7$ yr there is a general downward trend, though the values diverge and fluctuate greatly before reaching steady-state at times greater than $10^7$ yr. The enhancement at $\sim 10^6$ yr is due mainly to faster gas-phase destruction of key $DCO^+$ precursor species such as D and $CH_4D^+$ until the abundance of CO begins to drop at ca. $10^6$ yr, at which point production rates for $HCO^+$ begin to decline as well.

From Fig. 1(b) one can see that there clearly exists a correlation between the time-dependent behavior of the $H_2$ OPR and the cosmic ray ionization rate. Specifically, the higher the value of $\zeta$, the shorter the time needed for the OPR to reach steady-state. However, in order to determine the individual role of the initial OPR and cosmic ray ionization rate on the time-dependent $[HCO^+]/[DCO^+]$ abundance ratio, two more sets of 500 simulations were carried out in which only one of these parameters was varied at a time. Results from these simulations are shown in Fig. 2. From these figures, one can see that variations in the cosmic ray ionization rate and the OPR affect $R_D$ more strongly as time increases and have a pronounced effect on $R_D$ at times starting at ca. $10^5$ yr. Finally, at times considerably greater, only the variation of $\zeta$ has a significant effect on $R_D$. Thus, one can see that nuclear spin plays an important role in determining the $[HCO^+]/[DCO^+]$ abundance ratio through intermediate times, *but not thereafter*.

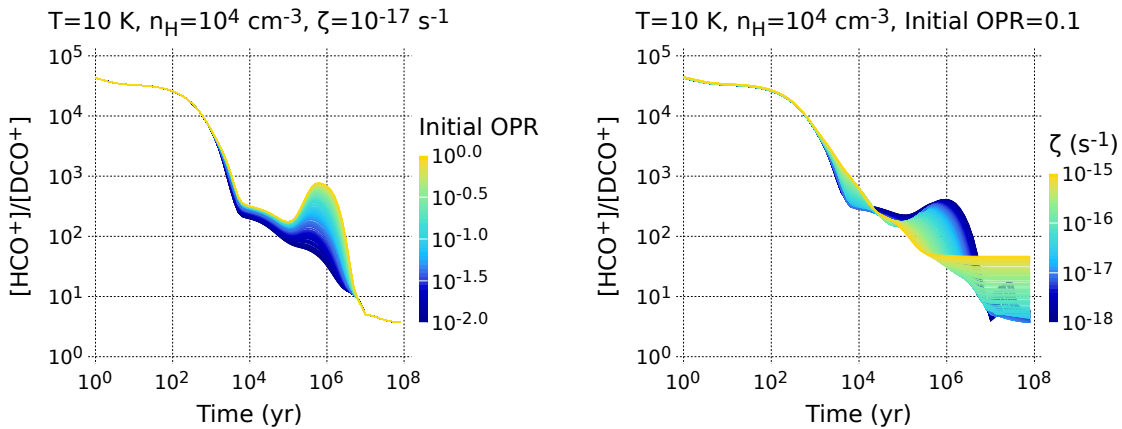

(a) Sensitivity analysis varying the initial OPR only

(b) Sensitivity analysis varying $\zeta$ only

Fig. 2.— Results from sensitivity analyses when only one parameter at a time is varied.

## 3.2. Time and Density Dependence

These results show that varying $\zeta$ and the initial OPR has a pronounced effect on $R_D$; however, whether and to what degree these effects change the ability to determine $\zeta$ from the observed value of $R_D$ remains unclear. Therefore, a subset of $R_D$ ratios at model times of $t = 10^5$, $10^6$, and $10^7$ yr was extracted from the dataset shown in Fig. 1(a) and plotted as a function of the cosmic ray ionization rate. Fig. 3 shows these abundance ratios, on top of which are superimposed data from additional sets of simulations under the same conditions but with a fixed initial OPR.

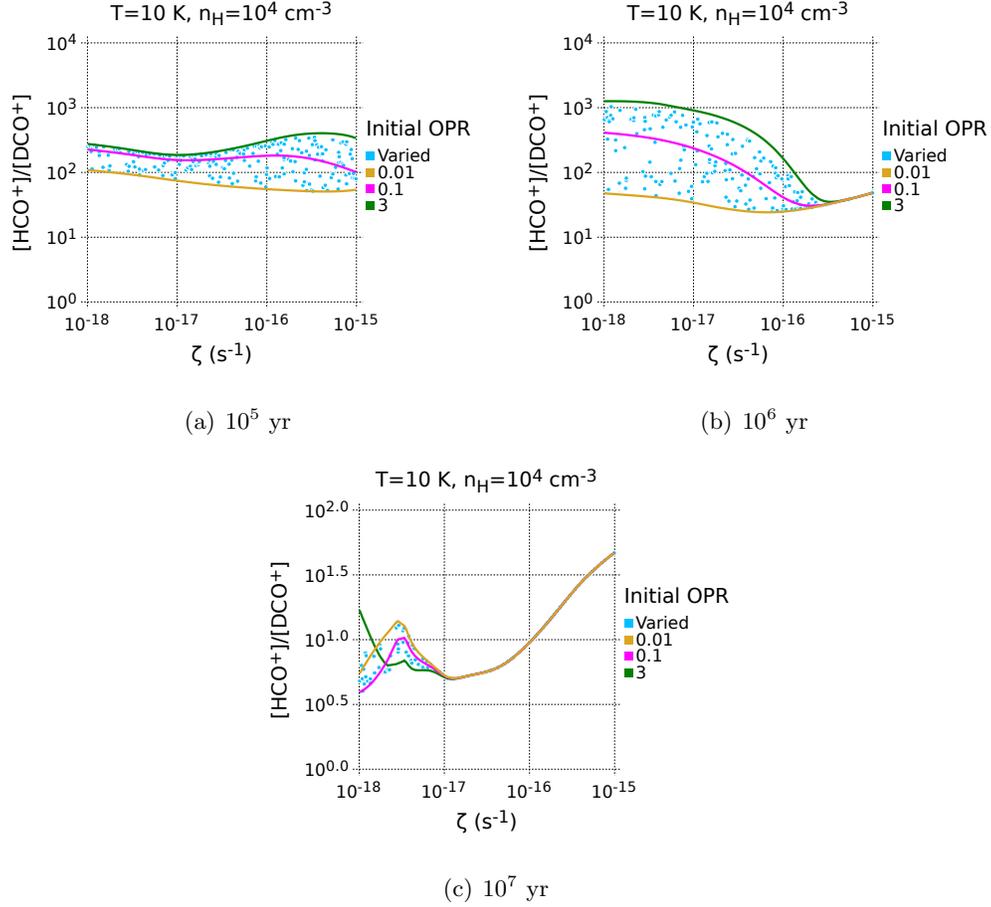

Fig. 3.— The [HCO$^+$]/[DCO$^+$] ratio vs. $\zeta$ (varied between between $10^{-18}$ and $10^{-15}$ s$^{-1}$) for a typical cold core at $10^5$ yr (panel a), $10^6$ yr (b), and $10^7$ yr (c). Shown here are data from models in which the initial OPR was varied randomly (blue) and those for which the initial OPR was set to 0.01 (yellow), 0.1 (magenta), and 3 (green).

From Fig. 3, one can see more clearly the influence of nuclear spin on the abundance ratio as a function of $\zeta$. When the initial OPR is kept fixed, the resultant data show that each ionization rate is correlated with a single, though not necessarily unique, abundance ratio. Nevertheless, at model times of $t < 10^7$ yr and if the initial OPR is sufficiently low, there will be very little difference in $R_D$ over the range of $\zeta$ between $10^{-18}$ and $10^{-15}$ s$^{-1}$, and so the use of $R_D$ as a tracer of $\zeta$ will not be robust. When the initial OPR is varied, two regimes become apparent. In one, the correlation breaks down and a region of uncertainty emerges, characterized by the spread of values due to variation of the initial OPR in the models. Moreover, the value of $\zeta$ that marks the transition between the correlated and uncorrelated regimes is highly time-dependent, occurring at $\zeta > 10^{-15}$ (not shown), $\zeta \approx 10^{-16}$, and $\zeta \approx 10^{-17}$ s$^{-1}$ for source ages of $10^5$, $10^6$, and $10^7$ yr, respectively. Based on these data, it appears that, for cold cores, at times $\geqslant 10^6$ yr, the correlation between $\zeta$ and the observed [HCO$^+$]/[DCO$^+$] ratios emerges and determinations of the former from the latter become safer, though again, the exact value of the cosmic ray ionization rate associated with the transition point from one regime to another depends on the source age.

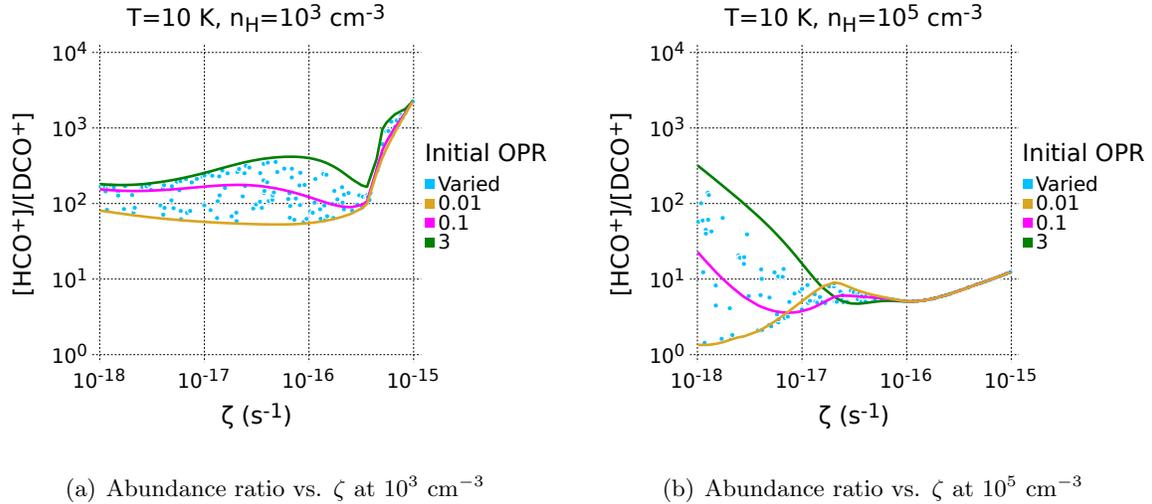

(a) Abundance ratio vs. $\zeta$ at $10^3$ cm$^{-3}$        (b) Abundance ratio vs. $\zeta$ at $10^5$ cm$^{-3}$

Fig. 4.— The [HCO$^+$]/[DCO$^+$] ratio vs. $\zeta$ at $10^6$ yr for a 10 K source with densities of $10^3$ (panel a) and $10^5$ (panel b) cm$^{-3}$. Shown are data from simulations in which the initial OPR is varied randomly and data in which it starts off at specific values.

As well as being time-dependent, the [HCO$^+$]/[DCO$^+$] probe of $\zeta$ in cold cores was found to be dependent on the source density. Shown in Fig. 4 are data from two sets of simulations which are plotted for densities higher and lower than our standard cold core by an order of magnitude. For both sets of simulations, the cosmic ray ionization rate was varied between $10^{-18}$ and $10^{-15}$ s$^{-1}$, and the initial OPR between 0.01 and 3. Results from the low density simulations at $10^6$ yr, given in Fig. 4(a), show that the uncorrelated regime extends roughly to $\zeta = 4 \times 10^{-16}$ s$^{-1}$, a higher transition value than for a $10^4$ cm$^{-3}$ density source at the same time. For ionization rates greater than this transition value, the [HCO$^+$]/[DCO$^+$] abundance ratio shows a steep increase in response to changes in $\zeta$. On the other hand, results from the high density simulations at $10^6$ yr, given in Fig. 4(b), show that the transition from one regime to another occurs near $\zeta \approx 5 \times 10^{-17}$ s$^{-1}$, roughly the standard value. Nevertheless, even in the high density - high $\zeta$ model where the correlation between the abundance ratio and ionization rate appears most robust, inferences of the latter from the former are complicated, both because of the generally slow change of the ratio to changes in the ionization rate for enhanced values of $\zeta$ and by uncertainties introduced from differences in the initial OPR of the source at low $\zeta$.

### 3.3. Results for SNR W51C

We next extend our evaluation to a real astrophysical source, W51C, an SNR interacting with a neighboring molecular cloud, which has been previously studied to infer $\zeta$ from the $[HCO^+]/[DCO^+]$ ratio (Ceccarelli et al. 2011). Bright $\gamma$-ray emission has been detected in this region, which is thought to arise from the interaction of cosmic rays from the SNR with the molecular cloud (Abdo et al. 2009). This cloud differs from cold cores in that it likely has an enhanced ionization rate and a gas temperature of 24 K with a density of $10^4$ cm$^{-3}$. There is also evidence that the cloud has two phases - a so-called high-ionization phase and a low-ionization phase (Ceccarelli et al. 2011). This result has not been obtained with our gas-grain code, as discussed in Section 4.1.

As before, sets of simulations of $R_D$ vs time were run both varying initial OPR values in the range 0.01 - 3 and using initial OPR values of 0.01, 0.1, and 3. Also as before, for all sets of simulations, $\zeta$ was varied between $10^{-18}$ and $10^{-14}$ s$^{-1}$. These data are shown in Fig. 5 for $10^5$ yr, $10^6$ yr, and $10^7$ yr along with the observed values of $R_D$ by Ceccarelli et al. (2011). As in the case for the cold core models, two regimes are clearly visible: before the OPR reaches steady state, a clear correlation between $\zeta$ and $R_D$ cannot be seen; however, at $10^6$ and $10^7$ yr, the variation in the abundance ratio caused by differences in the initial OPR disappear as these values converge at $\zeta \gtrsim 10^{-16}$ and $\zeta \gtrsim 10^{-17}$, respectively. The abundance ratios at $10^5$ yr converge as well, though at extremely high values of $\zeta \gtrsim 10^{-15}$ s$^{-1}$. As seen before in the cold core models, when the initial $H_2$ OPR is fixed, clear correlations between $[HCO^+]/[DCO^+]$ and $\zeta$ are recovered, and each ionization rate corresponds to a single abundance ratio. However, even when the initial OPR starts off at specific values, the slope of the abundance ratio vs $\zeta$ can be very shallow, especially at low to moderately enhanced ionization rates, and it is unclear that the value of $\zeta$ could be derived from observations of $[HCO^+]/[DCO^+]$ in this limit.

Because the cloud interacting with W51C is warmer than a typical cold core, model data for this environment may help elucidate the temperature dependence of $[HCO^+]/[DCO^+]$ as a probe of $\zeta$ when one considers the additional effects of nuclear spin. As one can see from comparing Fig. 3 with Fig. 5, the transition between the uncorrelated and correlated regimes occurs at approximately the same values of $\zeta$ for each of the model times shown, though predicted abundance ratios in the correlated regime are lower for the cold core than for the warmer source in W51C. Another key difference between the 10 K and 24 K models is that the degree of uncertainty is much greater in the former. In the 24 K W51C simulations where the initial OPR was varied, values for the abundance ratios are spread in the uncorrelated regime by less than an order of magnitude; however, for the 10 K cold core simulations, the spread of values, particularly around $10^6$ yr, is nearly two orders of magnitude.

It can be seen in Fig. 5 that for each of the times shown, the overlap between the observed $R_D$ value with uncertainty and the calculated values occurs in or near the correlated region, so the determination of $\zeta$ is secure, especially at the longer times. Comparison with the earlier results of Ceccarelli et al. (2011) is made in Section 4.

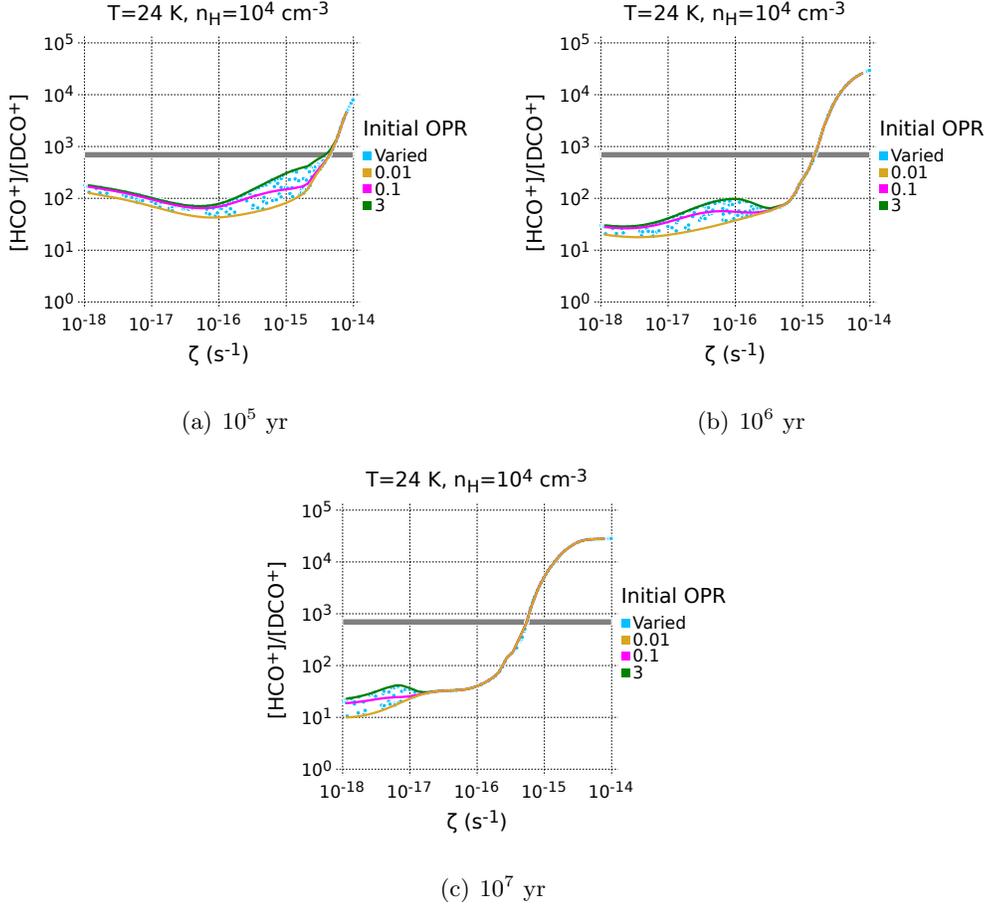

(a) $10^5$ yr

(b) $10^6$ yr

(c) $10^7$ yr

Fig. 5.— The [HCO$^+$]/[DCO$^+$] ratio at $10^5$ yr (panel a), $10^6$ yr (b), and $10^7$ yr (c) plotted as a function of $\zeta$ under the conditions of the molecular cloud in W51C. Pictured here are results for both sensitivity analyses in which the initial OPR starts off at specific values (lines) and is randomly varied (circles). The gray bar represents the range of observational abundance ratios from Ceccarelli et al. (2011).

### 3.4.  Results for L1174

L1174 is a dense core with an embedded star observed by Butner et al. (1995) to have an [HCO$^+$]/[DCO$^+$] ratio of $\sim 30$ and later modeled by Caselli et al. (1998) with the goal of inferring the cosmic ray ionization rate from $R_D$. From their observations, Butner et al. (1995) estimated L1174 to have a kinetic temperature of 15 K and density of $\sim 10^4$ cm$^{-3}$. As L1174 is similar to a cold core, we do expect large uncertainties in our approach except perhaps at long times. In order to further compare our models with these previous observational and theoretical results, we simulated L1174 using our chemical network, varying the initial OPR between 0.01 and 3 as before. Shown in Fig. 6 are the resulting computed values of $R_D$ at times of $10^5$, $10^6$, and $10^7$ yr, along with the observational values and uncertainties.

From the overlap of calculated and observed values of $R_D$, we can obtain the range of $\zeta$ although the uncertainties are large. At $10^5$ yr, however, there is no region of substantial overlap so the method fails. At $10^6$ yr, we obtain a large range for $\zeta$ of roughly $5 \times 10^{-18} - 10^{-15}$ s$^{-1}$ where there is a substantial overlap with observation. Finally, for $10^7$ yr, we reproduce the observed values at two very different but short ranges for $\zeta$: around $\sim 0.1$ and 100 times the standard value of $10^{-17}$ s$^{-1}$. These results are interpreted in Section 4.

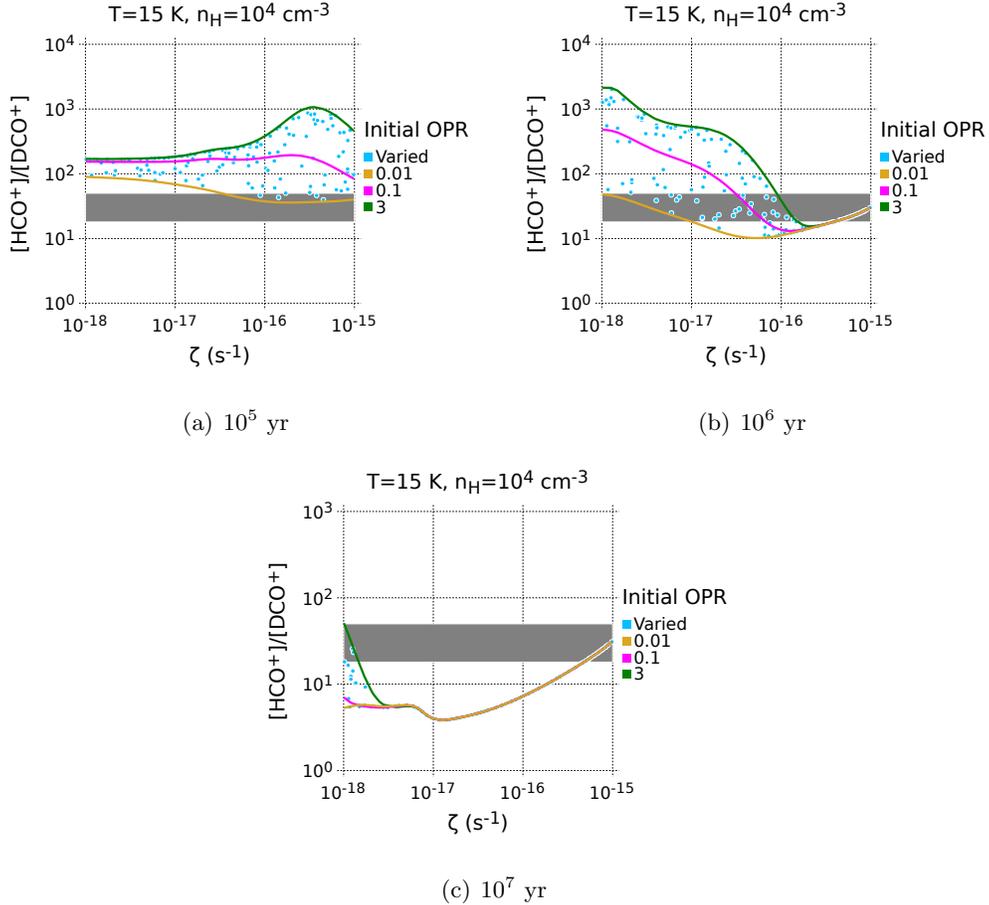

(a) $10^5$ yr

(b) $10^6$ yr

(c) $10^7$ yr

Fig. 6.— The [HCO$^+$]/[DCO$^+$] ratio at $10^5$ yr (panel a), $10^6$ yr (b), and $10^7$ yr (c) plotted as a function of $\zeta$ under the conditions of the dense core L1174. Pictured here are results for both sensitivity analyses in which the initial OPR starts off at specific values (lines) and varied randomly (circles). The gray bar represents the range of observational abundance ratios from Butner et al. (1995).

## 4. Discussion

In this work, we have reported the effects of a time-dependent and uncertain $OPR(H_2)$ on the use of the measured ratio $R_D = [HCO^+]/[DCO^+]$ in determining the cosmic ray ionization rate $\zeta$. Our simulations show that before the OPR reaches steady-state, different values of the OPR result in different values of $R_D$ for a given $\zeta$; this means that observations of $R_D$ cannot easily be used as a tracer of $\zeta$ in this regime. After the OPR reaches steady-state, there is a strong correlation between $[HCO^+]/[DCO^+]$ and $\zeta$. Furthermore, at higher temperatures, the regime in which $R_D$ and $\zeta$ are closely correlated appears earlier. Moreover, as $H_2$ density increases, faster total rates for reactions between $H_2$ and the ions $H^+$ and $H_3^+$ allow the OPR of the region to reach its steady-state value more quickly and extend the correlated regime over a wider range of ionization rates, as seen in Fig. 4. Thus, we suggest that it is more appropriate to use $[HCO^+]/[DCO^+]$ as a gauge for cosmic ray ionization in the limit of high densities and enhanced values of $\zeta$ since the OPR is more likely to be at or near steady-state.

A summary of the optimum conditions for using $[HCO^+]/[DCO^+]$ as a probe of $\zeta$ is given in Tables 1 and 2 for cold cores and the cloud in W51C, respectively. In these Tables, "..." represents conditions in which inference of $\zeta$ is not recommended, *unless there are some additional constraints on the initial OPR*, whereas +, ++, and +++ represent, from weakest to strongest robustness of correlation, regions in which such an inference may be possible, based on our results.

It should be noted that the technique used here for determining $\zeta$ based on $[HCO^+]/[DCO^+]$ abundances is not recommended for well studied sources for which a more detailed approach can be used. One such detailed approach is the combined chemical and radiative transfer method of Maret & Bergin (2007), who examined the prestellar core B68. This model, which used observations of $H^{13}CO^+$ and $DCO^+$, was able to put constraints not only on $\zeta$, but also on the metallicity and current OPR of the source. For well-studied objects like B68 for which there is sufficient data for chemical abundance and physical parameter profiles, a more accurate alternative approach like the one used by Maret & Bergin (2007) is desirable; however, this method may not be easy to apply to large numbers of less well understood sources. Therefore it is important to explore how much information one can get from a simplified analysis, as well as the accuracy and limitations of such a scheme. Thus, the main purpose of this paper is not to find the best method for constraining $\zeta$ in interstellar sources, rather, it is to explore the robustness of [HCO$^+$]/[DCO$^+$] as a probe for $\zeta$. As mentioned earlier, use of this abundance ratio to constrain the cosmic ray ionization rate has been previously suggested by Guelin et al. (1977) and others; however our focus concerns the constraints we have for other parameters, such as temperature and density, and especially on the effects of nuclear-spin chemistry. We further emphasize that the role of the time dependence of the OPR of H$_2$ has not been explored previously in this context.

Table 1. Optimum conditions for the robustness of $[HCO^+]/[DCO^+]$ as a probe of $\zeta$ for cold cores

| $\zeta$ | Early Time $10^5 - 10^6$ yr | Intermediate Time $10^6 - 10^7$ yr | Late Time $10^7 - 10^8$ yr |
|---|---|---|---|
| $\zeta < 1.3 \times 10^{-17}$ s$^{-1}$ | ... | ... | ... |
| $\zeta \approx 1.3 \times 10^{-17}$ s$^{-1}$ | ... | ... | ... |
| $\zeta > 1.3 \times 10^{-17}$ s$^{-1}$ | ... | ... | + |

## 4.1. Comparison with Earlier Results

One source where an estimation of $\zeta$ from the [HCO$^+$]/[DCO$^+$] ratio is likely safe is the molecular cloud near the supernova remnant W51C, and a comparison can be made between our results and earlier results using the same abundance ratio, as described in Ceccarelli et al. (2011). These authors estimated $\zeta$ to have a value of $1.0 - 1.3 \times 10^{-15}$ s$^{-1}$, roughly two orders of magnitude greater than the standard value. However, the chemical network used in that study was limited to purely gas-phase reactions and steady-state conditions (Le Petit et al. 2006). In addition, the effects of nuclear spin were not explicitly considered, though a constant OPR of less than 0.01 was assumed. Thus it was not possible to observe the uncertainties that result from variations in the initial OPR and their subsequent time dependence as studied in this work. In their study, Ceccarelli et al. (2011) further noted the appearance of both a high-ionization phase (HIP) and low-ionization phase (LIP), with a sharp transition point between the two occurring at ca. 120 times the standard ionization rate.

Our simulations of W51C show no evidence of these bistable phases and our results all pertain to a high-ionization region. On the other hand, we have observed, to the best of our knowledge, the first hint of bistability in a time-dependent gas-grain code in our L1174 simulations. There, a discontinuity appears in values of $R_D$ vs. $\zeta$ for times greater than $10^6$ yr. The discontinuity lies at even higher values of $\zeta$ than that reported by Ceccarelli et al. (2011) and is not shown in Fig. 6. At such values, it is uncertain whether our chemical network is accurate under the relevant physical conditions.

One uncertainty in our results has to do with the unknown chemical age of the molecular cloud near W51C. The supernova remnant itself is "middle-aged", which refers to an age of $\approx 3 \times 10^4$ yr (Koo et al. 1995). The molecular cloud is presumably older than this age, but at ages greater than that of the SNR and the $\gamma$-ray source, likely had a lower temperature. Thus it is best to consider the inferred value of $\zeta$ over a wide range of possible chemical lifetimes. Our inferred values of $\zeta$, obtained by comparison of the observed [HCO$^+$]/[DCO$^+$] ratio of $700 \pm 100$ (Ceccarelli et al. 2011) with our values for a wide variety of ages are presented in Table 3 for the three time eras $10^5$ - $10^6$ yr, $10^6$ - $10^7$ yr, and $10^7$ - $10^8$ yr. The uncertainties presented correspond to the use of ranges of times rather than specific times. The inferred values of $\zeta$ vary from $3 \times 10^{-15}$ s$^{-1}$ at the earliest times to almost an order of magnitude lower at the latest times, with a value of $1 \times 10^{-15}$ s$^{-1}$ at the intermediate range. This last value corresponds very well with that obtained by Ceccarelli et al. (2011), but even the largest discrepancy between our values and the earlier result is only a factor of three. Since, as shown in Figures 5(b) and 5(c), the observed abundance ratios map to ionization rates in the safer, correlated regime, our confidence in our results is increased.

Table 2. Optimum conditions for the robustness of [HCO$^+$]/[DCO$^+$] as a probe of $\zeta$ for W51C Cloud

| $\zeta$ | Early Time $10^5 - 10^6$ yr | Intermediate Time $10^6 - 10^7$ yr | Late Time $10^7 - 10^8$ yr |
|---|---|---|---|
| $\zeta < 1.3 \times 10^{-17}$ s$^{-1}$ | ... | ... | + |
| $\zeta \approx 1.3 \times 10^{-17}$ s$^{-1}$ | ... | + | ++ |
| $\zeta > 1.3 \times 10^{-17}$ s$^{-1}$ | + | ++ | +++ |

L1174 resembles to a certain extent the cold cores, the conditions of which for robustness of our approach are listed in Table 1. L1174 was modeled by Caselli et al. (1998), who used steady-state abundances from a purely gas-phase chemical network, and obtained a rather tight value for $\zeta$ of (1.3 - 2.6) $\times 10^{-16}$ s$^{-1}$. From Table 1 and previous discussions, we expect that our approach will not be as unambiguous. To determine what age would be best for this source, we can start with the isothermal free-fall collapse time using the formalism of Shu (1977), which predicts a dynamical timescale of a few times $10^5$ yr for dense cores to form starting from a translucent cloud with a density of $10^3$ cm$^{-3}$. If the progenitor cloud is diffuse ($10^2$ cm$^{-3}$), then the dynamical time is increased by roughly an order of magnitude to a few times $10^6$ yr. These times may thus be used as rough estimates for lower limits to the ages of such sources, especially for dense cores with embedded stars, since these will be older by some amount than the purely dynamical age. Since L1174 both has an embedded protostar and is thus older than the purely dynamical time and is not in the vicinity of a strong $\gamma$-ray source such as W51C, it seems reasonable that the results for $\zeta$ at $10^6$ or $10^7$ yr should pertain. At the earlier time, however, our approach does not yield a very precise result; as can be seen in Fig. 6 all we can state is that $\zeta$ lies in the broad range $5 \times 10^{-18} - 10^{-15}$ s$^{-1}$. If we arbitrarily limit our attention to the correlated region, the range is reduced to $10^{-16} - 10^{-15}$ s$^{-1}$, in better agreement with the result of Caselli et al. (1998). At $10^7$ yr, we obtain two discrete range of values, with the more definitive one, obtained in the correlated region, $5 \times 10^{-16} - 1 \times 10^{-15}$ s$^{-1}$. The other range of values lies at $\leq 0.1$ of the standard value of $\zeta$ and involves only a partial overlap of the observational and computed $R_D$ values. The upper range of values lies in good agreement with the steady-state result of Caselli et al. (1998), but is rather high for this type of source, which does not contain any obvious intense external irradiation.

Table 3.  Inferred cosmic ray ionization rates for the W51C Cloud

| $10^5 < t < 10^6$ yr | $10^6 < t < 10^7$ yr | $10^7 < t < 10^8$ yr |
|---|---|---|
| $(3.0 \pm 1.5) \times 10^{-15}$ s$^{-1}$ | $(1.0 \pm 0.5) \times 10^{-15}$ s$^{-1}$ | $(3.6 \pm 2.0) \times 10^{-16}$ s$^{-1}$ |

## 4.2. Effects of Additional Processes

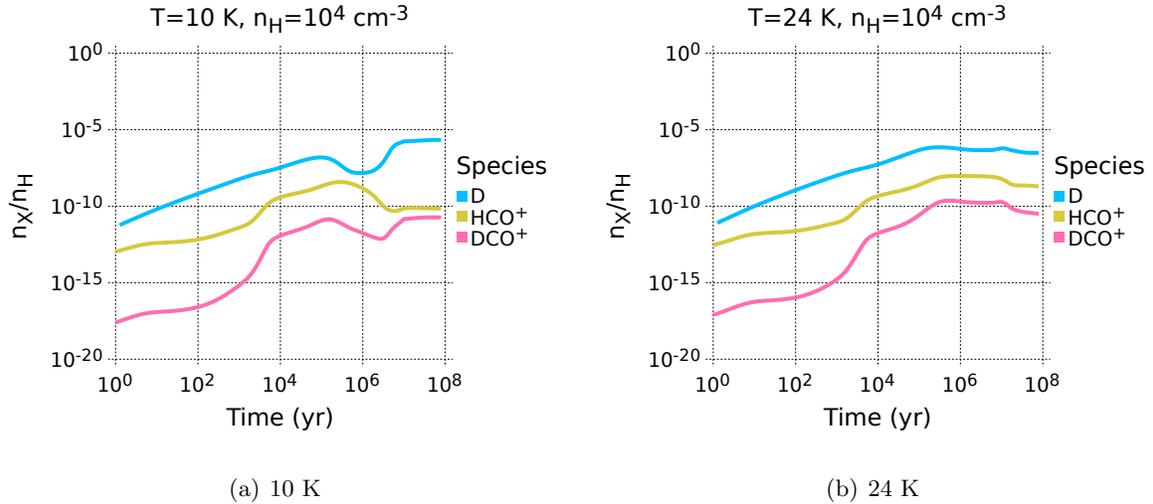

(a) 10 K                    (b) 24 K

Fig. 7.— Abundances relative to $n_H$ for D, $HCO^+$, and $DCO^+$ for simulations run at 10 K (a) and 24 K (b) using the standard ionization rate and an initial OPR of 3.

From a comparison of Fig. 3 and Fig. 5, especially the panels for a time of $10^6$ yr, temperature appears to have a pronounced effect on the size and behavior of the uncorrelated regime of $[HCO^+]/[DCO^+]$ values, which lie below the transition point to the correlated regime. An analysis of the total rates of reactions for the simulations shown in Fig. 3 and Fig. 5 points to one reaction in particular being behind these effects; namely,

$$HCO^+ + D \longrightarrow DCO^+ + H,$$ (8)

the total rate of which is roughly three orders of magnitude greater at 24 K than at 10 K. In these models, the atomic D is formed mainly via reaction (5), the dissociative recombination of $DCO^+$. The impact of reaction (8) on the $[HCO^+]/[DCO^+]$ ratio was previously considered by Dalgarno & Lepp (1984), who suggested its potential importance, even in dense regions. Moreover, Öberg et al. (2015) recently noted, in the context of protoplanetary disk chemistry, that the importance of reaction (8) as a formation pathway for $DCO^+$ is most prominent at elevated temperatures. Our results are in agreement with both studies.

Shown in Fig. 7 are relative abundances for $HCO^+$, $DCO^+$, and D for individual 10 K and 24 K simulations using the standard ionization rate and an initial OPR of 3. From these, one can see that the total rate for reaction (8) is increased because of the greatly enhanced gas-phase deuterium abundance in the 24 K model at intermediate times. A comparison of the total rates for both models indicates that deuterium desorption is the key process behind both the different gas-phase D abundances and the previously mentioned difference in $[HCO^+]/[DCO^+]$ values at 10 K and 24 K. Briefly, since the rate of deuterium desorption is much higher at elevated temperatures, the total rate of reaction (8) is greater, leading to an increased $DCO^+$ abundance and, by extension, a reduced $[HCO^+]/[DCO^+]$ value, compared with a 10 K region, especially at intermediate times. When alternate formation pathways dominate, such as the one given in reaction (8) for $DCO^+$ at 24 K, the spread of $[HCO^+]/[DCO^+]$ values, caused mainly by the importance of the pathways given in reactions (1) and (2), is reduced and effects caused by differences in the initial OPR are minimized. Thus, the uncorrelated region, in which $\zeta$ cannot easily be inferred from $[HCO^+]/[DCO^+]$, is minimized at intermediate times at elevated temperatures.

### 4.2.1. Surface Interconversion of ortho and para Hydrogen

Recently, Bron et al. (2016) have shown the potential importance of surface-facilitated interconversion of ortho and para $H_2$. In this study, the authors simulated the ortho-para surface-interconversion of $H_2$ using the current state of experimental data, and found that it can occur with a high efficiency. Surface conversion of ortho and para hydrogen was not included in the present work, however, since the process is still not fully understood. The parameters used in determining the efficiency of the process are often ill-constrained, specifically the $H_2$ desorption energies, $E_D$ (Fukutani & Sugimoto 2013). The binding energies, in turn, influence the dust temperature region of maximum efficiency. Bron et al. (2016) simulated surface ortho-para conversion using a range of desorption energies for $H_2$ and found that the process is only efficient over a dust temperature region of 7-26 K, outside of which the surface conversion efficiency falls rapidly to zero. Thus, depending on the interstellar source environment, surface conversion may be critical or unimportant for fully understanding the evolution of the OPR with time. Assuming surface conversion is important, its effect on these results would be to reduce the time to reach steady-state OPR values, thereby shrinking the regime of uncertainty and making inferences of $\zeta$ from $[HCO^+]/[DCO^+]$ more justified for lower values of $\zeta$ and lower times. In addition, given the high degree of temperature dependence of the process, it is more likely to be important in cold cores. In the case of W51C, the potential effects of surface conversion are more likely to be negligible, since its warmer temperature of 24 K puts it near the cut-off point for efficient ortho-para conversion of $H_2$.

## 5. Conclusions

The main conclusions of this work are as follows:

1. The initial OPR of $H_2$ in a source can influence the $[HCO^+]/[DCO^+]$ abundance ratio, partic-

ularly when $\zeta < 10^{-17} s^{-1}$, having the effect of breaking down the close correlation between the abundance ratio and $\zeta$.

2. The breakdown in the correlation between $\zeta$ and $[HCO^+]/[DCO^+]$ is time-dependent, and depends on the unique physical conditions of the source, such as $n_H$ and temperature.

3. The correlation between $\zeta$ and $[HCO^+]/[DCO^+]$ is most robust in old sources and those with high ionization rates, high temperatures, and large densities.

4. Our range of inferred values for $\zeta$ for the molecular cloud near the supernova remnant W51C lies in the range $4 \times 10^{-16}$ s$^{-1}$ - $3 \times 10^{-15}$ s$^{-1}$ with the major uncertainty the age of the cloud, and secondarily the history of the cloud temperature. Our values all lie in reasonable agreement with the earlier result of Ceccarelli et al. (2011); the best agreement occurs during a cloud age of $10^6$ - $10^7$ yr.

5. For the dense core L1174, our approach is not able to obtain a well determined robust result for $\zeta$ at either $10^5$ yr or $10^6$ yr. Two results are obtained at a longer age of $10^7$ yr, with the larger range of values for $\zeta$, which lie in the correlated region, more robust. This range of values is in reasonable agreement with the steady-state results of Caselli et al. (1998) although both are very large for a source, such as L1174, without an external source of intense irradiation.

EH wishes to thank the National Science Foundation for continuing to support the astro-chemistry program at the University of Virginia. He thanks as well the NASA Exobiology and Evolutionary Biology Program through a subcontract from Rensselaer Polytechnic Institute. JBB acknowledges funding from an NSF Graduate Research Fellowship.